\documentclass[journal]{IEEEtran}

\usepackage{authblk}

\usepackage{framed,multirow}

\usepackage{amssymb}
\usepackage{latexsym}

\usepackage{url}
\usepackage{xcolor}

\usepackage{amsmath}
\usepackage{graphicx}
\usepackage{caption}
\usepackage{subcaption}
\usepackage{wrapfig}
\usepackage{float}
\usepackage{amsthm}
\usepackage{diagbox}
\usepackage{stmaryrd}
\usepackage{comment}
\usepackage{color,soul}
\usepackage{multirow}

\usepackage{makecell}

\usepackage{xcolor,cancel}

\setstcolor{red}

\floatstyle{plaintop}
\restylefloat{table}

\usepackage{hyperref}
\hypersetup{
	colorlinks=true,
	linkcolor=blue,
	filecolor=magenta,      
	urlcolor=cyan,
}

\begin{document}

\title{ProsRegNet: A Deep Learning Framework for Registration of MRI and Histopathology Images of the Prostate}

\author[1]{Wei Shao}
\author[2]{Linda Banh}
\author[3]{Christian A. Kunder}
\author[4]{Richard E. Fan}
\author[4]{Simon J. C. Soerensen}
\author[5]{Jeffrey B. Wang}
\author[4]{Nikola C. Teslovich}
\author[6]{Nikhil Madhuripan}
\author[7]{Anugayathri Jawahar}
\author[1]{Pejman Ghanouni}
\author[4]{James D. Brooks}
\author[1,4]{Geoffrey A. Sonn}
\author[1]{Mirabela Rusu}

\affil[1]{Department of Radiology, Stanford University, Stanford, CA 94305, USA}
\affil[2]{Department of Electrical Engineering, Stanford University, Stanford, CA 94305, USA}
\affil[3]{Department of Pathology, Stanford University, Stanford, CA 94305, USA}
\affil[4]{Department of Urology, Stanford University, Stanford, CA 94305, USA}
\affil[5]{School of Medicine, Stanford University, Stanford, CA 94305, USA}
\affil[6]{Department of Radiology, University of Colorado, Aurora, CO 80045, USA}
\affil[7]{Loyola University Medical Center, Maywood, IL 60153, USA}


\maketitle

\begin{abstract}
Magnetic resonance imaging (MRI) is an increasingly important tool for the diagnosis and treatment of prostate cancer. 
However, interpretation of MRI suffers from high inter-observer variability across radiologists, thereby contributing to missed clinically significant cancers, overdiagnosed low-risk cancers, and frequent false positives. 
Interpretation of MRI could be greatly improved by providing radiologists with an answer key that clearly shows cancer locations on MRI.
Registration of histopathology images from patients who had radical prostatectomy to pre-operative MRI allows such mapping of ground truth cancer labels onto MRI. 
However, traditional MRI-histopathology registration approaches are computationally expensive and require careful choices of the cost function and registration hyperparameters. 
This paper presents ProsRegNet, a deep learning-based pipeline to accelerate and simplify MRI-histopathology image registration in prostate cancer. 
Our pipeline consists of image preprocessing, estimation of affine and deformable transformations by deep neural networks, and mapping cancer labels from histopathology images onto MRI using estimated transformations.
We trained our neural network using MR and histopathology images of 99 patients from our internal cohort (Cohort 1) and evaluated its performance using 53 patients from  three different cohorts (an additional 12 from Cohort 1 and 41 from two public cohorts).
Results show that our deep learning pipeline has achieved more accurate registration results and is at least 20 times faster than a state-of-the-art registration algorithm. 
This important advance will provide radiologists with highly accurate prostate MRI answer keys, thereby facilitating improvements in the detection of prostate cancer on MRI.
Our code is freely available at \url{https://github.com/pimed//ProsRegNet}.
\end{abstract}



\section{Introduction}
Prostate cancer is the second leading cause of cancer death and the most diagnosed cancer in men in the United States, with an estimated 33,330 deaths and 191,930 new cases in 2020~\cite{prostate_cancer_stats}.
Diagnosis, staging, and treatment planning of prostate cancer is increasingly assisted by magnetic resonance imaging (MRI)~\cite{TurkbeyL.Baris2012MMap,VermaSadhna2012Oodc}.
The Prostate Imaging Reporting and Data System (PI-RADS)~\cite{WeinrebJeffreyC2016PPI} was developed to standardize the acquisition, interpretation, and reporting of prostate MRI.
Despite the widespread adoption of PI-RADS, the performance of MRI still suffers from high levels of variation across radiologists~\cite{SonnGeoffreyA2019PMRI}, reduced  positive predictive value (27-58\%)~\cite{WestPhalen_2020}, low inter-reader agreement (k = 0.46-0.78)~\cite{AhmedHashimU2017Daom},  and large variations in reported sensitivity (58-96\%) and specificity (23-87\%)~\cite{AhmedHashimU2017Daom}.
It also has been shown that high interobserver disagreement in prostate MRI significantly affects prostate biopsy practice including aborting planned biopsy and reduced number of region of interest samples~\cite{Rosenzweig_2020}.
One major barrier to improvement in MRI interpretation is the lack of a pathologic reference standard to provide radiologists detailed feedback about their performance. 
Image registration~\cite{Shao_2016_CVPR_Workshops} of the pre-surgical MRI with histopathology images after surgical resection of the prostate (radical prostatectomy) addresses this issue by enabling mapping of the extent of cancer from the ground-truth histopathology images onto the MRI. 
Such mapping allows side-by-side comparison of the histopathology and MRI images, which can be use in the training of radiologists to improve their interpretation of MRI.
Furthermore, accurate cancer labels achieved by image registration may facilitate the development of radiomic and deep learning approaches for early prostate cancer detection and risk stratification on pre-operative MRI~\cite{CaoRuiming2019JPCD,correlation,ZhiweiWang2018ADoC,bhattacharya2020corrsignet}.

Several MRI-histopathology image registration approaches have been developed to account for elastic tissue deformation occurred during histological preparation inducing tissue fixation, sectioning, and mounting on histologic slides.
Turkbey et al. developed patient-specific 3D printed molds for the resected prostate that are designed based on the pre-operative MRI and allow sectioning of the prostate in-plane with the same slice thickness as MRI~\cite{TurkbeyBaris2011M3PM}.
A radiologist will first carefully segment the edge of the prostate gland as well as indicate areas of suspected prostate cancer. 
	From this segmentation, a 3D volume will be created, which is then imported into a modeling software to create a personalized mold such that the orientation of the prostate specimen is aligned with the original MRI to guide the gross sectioning of the ex vivo prostate to have exact slice correspondences with the MRI~\cite{priester2014system}.
Several approaches~\cite{LOSNEGARD201824,Rusu2019Registration,WuHoldenH.2019Asup} rely on patient-specific 3D printed molds to establish better histopathology and MRI slice correspondences in order to improve the registration of MRI and histopathology images.
While some approaches work directly with MR and histopathology images alone, others require additional steps including a separate \textit{ex vivo} MRI of the prostate~\cite{WuHoldenH.2019Asup}, fiducial markers~\cite{WardAaronD2012Prod}, or advanced image similarity metrics~\cite{ChappelowJonathan2011Erom,LiLin2017Coev}.
Several pipelines have been developed for direct integration of MR and histopathology images by 3D histopathology volume reconstruction~\cite{LOSNEGARD201824,Rusu2019Registration,SamavatiNavid2011Bmdr,StilleMaik20133ro2}, but most are time-consuming, computationally expensive, and can suffer from partial volume effects and artifacts due to large spacing between images.

Typically, a geometric transformation can be parameterized by either a few (affine) or a large number of (deformable) variables.
Previous automated MRI-histopathology registration approaches estimate variables that encode geometric transformations by optimizing a cost function for tens or hundreds of iterations~\cite{GoubranMaged2013Iroe,GoubranMaged2015Roit,RusuMirabela2017CopC}.
Therefore, this optimization process is computationally expensive and can take several minutes to finish.
Moreover, the estimated transformation is often sensitive to the choice of hyperparameters (e.g., the number of iterations and the cost function), making traditional registration approaches complex to set up and reducing their generalization.

To address this important gap, this paper presents a deep learning based pipeline for efficient MRI-histopathology registration.
In the past few years, deep learning has been successfully used in many medical image registration problems.
A deep learning based registration network can be considered as a function that takes two images, a fixed image and a moving image, as the input and directly outputs a unique transformation without requiring additional optimization.
Many deep learning approaches~\cite{balakrishnan2018unsupervised,BalakrishnanGuha2019VALF,dalca2018unsupervised,GhosalSayan2017DdrE,krebs2019learning, YangXiao2017QFpi,ZhangJun2018IDNf} assume that the fixed and moving images have already been aligned by affine registration and only focus on the deformable registration. However, the affine registration of MRI and histopathology images of the prostate is challenging since they are considerably different modalities while having different contents. Therefore, prior deformable registration approaches cannot be directly used for MRI histopathology registration where affine registration is a necessity due to large geometric changes of the prostate during histological preparation. Rocco et al. proposed a multi-stage deep learning framework (CNNGeometric) that can handle both affine and deformable deformations of natural images~\cite{Rocco17}.
Inspired by their study, we developed the ProsRegNet registration pipeline for affine and deformable registration of the MRI and histopathology images.
Our registration pipeline includes preprocessing and postprocessing modules, and the registration network that estimates an affine transformation at the first stage and a more accurate thin-plate-spline transformation at the second stage.
Some other deep learning registration approaches can also jointly estimate the affine and deformable transformations~\cite{de2019deep,shen2019networks}.
Similar to our ProsRegNet approach, the approach developed in~\cite{de2019deep} used a feature extraction network followed by a parameter estimation network. Unlike our approach, their model lacked the feature matching component which has been shown to increase the generalization capabilities of registration networks to unseen images~\cite{Rocco17}.
In our study, we will show that our ProsRegNet registration network trained with images from one cohort generalizes well to unseen images from other cohorts. 
Moreover, the models developed by~\cite{de2019deep,shen2019networks} were trained in an unsupervised manner using the normalized cross correlation, which can be unsuitable for MRI-histopathology registration as the intensities are not correlated. To our knowledge, we are the first to
apply deep learning to the problem of MRI-histopathology registration of the prostate.
We will demonstrate that our deep learning registration pipeline can achieve better registration accuracy than the state-of-the-art RAPSODI approach~\cite{rusu2020registration} while being much more computationally efficient and  easier to use for non-experts users.

This paper has the following major contributions:
\begin{itemize}
	\item We are the first to use deep learning to solve the challenging problem of registering MRI and histopathology images of the prostate.
	\item We avoid the shortcomings of multi-modal similarity measures for MRI-histopathology registration by training our registration network with mono-modal synthetic image pairs in an unsupervised manner using a mono-modal dissimilarity measure. During the testing, we applied our network to multi-modal image registration as the network has learned how to solve image registration problems irrespective of the image modalities. 
	\item We improved the stability of the training by parameterizing the transformations using the sum of an identity transform and the estimated parameter vector scaled by a small weight.
	\item We trained our network with a large set of MRI and histopathology prostate images and evaluated our approach relative to the state-of-the-art traditional and deep learning registration methods.
	\item Our code is one of the very few freely available MRI-histopathology registration codes.
\end{itemize}

\section{Materials and Methods}

\subsection{Data Acquisition}
This study approved by the Institutional Review Board at Stanford University included 152 subjects with biopsy-confirmed prostate cancer from three cohorts at different institutions. 
The first cohort consists of 111 patients who had a pre-operative MRI scan and underwent radical prostatectomy at Stanford University.
The excised prostate was submitted for histological preparation and we used a patient-specific 3D printed mold to generate whole-mount histopathology images that had slice-to-slice correspondences with the MRI.
Experts determined the correspondences between T2-weighted (T2-w) MRI and histopathology slices.
The prostate, cancer, urethra, and other anatomic landmarks on histopathology images were manually segmented  by an expert genitourinary pathologist.
Two hundred fifty-seven anatomic landmarks visible on both MRI and histopathology images, e.g., benign prostate hyperplasia nodules and ejaculatory ducts were chosen for a subset of 12 subjects from the first cohort.
The second cohort consisted of 16 patients from the publicly available ``Prostate Fused-MRI-Pathology" dataset in The Cancer Imaging Archive (TCIA) [dataset]~\cite{MadabhushiAnant2016FRPD}.
Each patient had an MRI along with digitized histopathology images of the corresponding radical prostatectomy specimen. 
Each surgically excised prostate specimen was originally sectioned and quartered resulting in four images for each section. 
The four images were then digitally stitched together to produce a pseudo-whole mount section.
Annotations of cancer presence on the pseudo-whole mount sections were made by an expert pathologist.
Slice correspondences were established between the individual T2-w MRI and stitched pseudo-whole mount sections by the program in~\cite{TothRobert2014HAis} and checked for accuracy by an expert pathologist and radiologist.
The third cohort consisted of 25 patients from the publicly available TCIA ``Prostate-MRI" dataset [dataset]~\cite{ChoykePeter2016DFP}.
Each patient had a pre-operative MRI and underwent a radical prostatectomy. 
A mold was generated from each MRI, and the prostatectomy specimen was first placed in the mold, then cut in the same plane as the MRI. 
The data was generated at the National Cancer Institute, Bethesda, Maryland, USA between 2008-2010.
For all of the three cohorts, the prostate on each MRI scan was manually segmented and used in the registration procedure.
The prostate segmentation serves to drive the alignment while the urethra and other anatomic landmarks were only used to evaluate the registration.
We summarized details of datasets from the above three cohorts in Table~\ref{table:data_summary}.
\begin{table*}
	\small
	\caption{Summary of datasets. We used 152 = 111 + 16 + 25 patients from three cohorts. T2-w MRI: T2-weighted MRI, 
		H\&E: Hematoxylin and Eosin, TR:  repetition time, TE: echo time, H: in-plane image height, W: in-plane image width, D: through-plane image depth.}
	\begin{tabular}{|c|c|c|c|c|c|c|}
		\hline
		& \multicolumn{2}{c|}{Cohort 1 (Stanford)} & \multicolumn{2}{c|}{Cohort 2 (TCIA)} & \multicolumn{2}{c|}{Cohort 3 (TCIA)} \\ \hline
		\begin{tabular}[c]{@{}c@{}}Number of \\ patients\end{tabular} & \multicolumn{2}{c|}{111} & \multicolumn{2}{c|}{16} & \multicolumn{2}{c|}{25} \\ \hline
		Modality & MRI & Histology & MRI & Histology & MRI & Histology \\ \hline
		Manufacturer & GE & - & Siemens & - & Philips & - \\ \hline
		Coil type & Surface & - & Endorectal & - & Endorectal & - \\ \hline
		Sequence & T2-w MRI & Whole-mount & T2-w MRI & \begin{tabular}[c]{@{}c@{}}Pseudo-whole\\ mount\end{tabular} & T2-w MRI & \begin{tabular}[c]{@{}c@{}}Blockface-whole\\ mount\end{tabular} \\ \hline
		\begin{tabular}[c]{@{}c@{}}Acquisition \\ characteristics\end{tabular} & \begin{tabular}[c]{@{}c@{}}TR: {[}3.9s, 6.3s{]}, \\ TE: {[}122ms, 130ms{]}\end{tabular} & \begin{tabular}[c]{@{}c@{}}H\&E stained, \\ 3D-printed mold\end{tabular} & \begin{tabular}[c]{@{}c@{}}TR: {[}3.7s, 7.0s{]},\\ TE: 107ms\end{tabular} & H\&E stained & \begin{tabular}[c]{@{}c@{}}TR: 8.9s,\\ TE: 120 ms\end{tabular} &\begin{tabular}[c]{@{}c@{}} H\&E stained, \\ mold \end{tabular} \\ \hline
		Image size & \begin{tabular}[c]{@{}c@{}}H,W: \{256,512\},\\ D: {[}24,43{]}\end{tabular} & H,W:{[}1663,7556{]} & \begin{tabular}[c]{@{}c@{}}H,W: 320x320,\\ D: {[}21,31{]}\end{tabular} & H,W:{[}2368, 6324{]} & \begin{tabular}[c]{@{}c@{}}H,W: 512,\\ D: 26\end{tabular} & H,W: {[}496,2881{]} \\ \hline
		\begin{tabular}[c]{@{}c@{}}In-plane \\ resolution (mm)\end{tabular} & {[}0.27, 0.94{]} & \{0.0081,0.0162\} & {[}0.41,0.43{]} & 0.0072 & 0.27 & \{0.0846,0.0216\} \\ \hline
		\begin{tabular}[c]{@{}c@{}}Distance \\ between slices\end{tabular} & {[}3mm,5.2mm{]} & {[}3mm,5.2mm{]} & 4mm & Free hand & 3mm &  3mm \\ \hline
	\end{tabular}
	\label{table:data_summary}
\end{table*}

\subsection{State-of-the-art RAPSODI Registration Framework}
We briefly summarize the state-of-the-art RAPSODI (\underline{Ra}diology \underline{p}athology \underline{s}patial \underline{o}pen-source multi-\underline{d}imensional \underline{i}ntegration) framework for the registration of MRI and histopathology images~\cite{rusu2020registration}.
The RAPSODI approach assumes known slice correspondences between MRI and histopathology images, and starts with 3D reconstruction of the histopathology specimen by registering each histopathology slice to its adjacent slice.
The purpose of the 3D reconstruction of the histopathology volume is to initialize the histopathology slices in the registration with the MRI.
Then 2D rigid, affine and deformable transformations between each histopathology image and the corresponding T2-w MRI slice are estimated iteratively using gradient descent.
The rigid and affine registrations use the prostate masks as the input and the sum of squared differences as the cost function.
The deformable registration uses the images masked by the prostate segmentation as the input, free-from deformations as the deformation model and the Mattes mutual information as the cost function.
Early stopping is used in the deformable registration to prevent overfitting.
Compared to our deep learning registration approach, RAPSODI requires significant user input including careful choice of similarity metric and registration hyperparameters such as step size, and the number of iterations.
The RAPSODI approach has been shown to be highly accurate and we will compare it with our deep learning registration pipeline.

\subsection{Deep Learning ProsRegNet Pipeline}
We propose the ProsRegNet (\underline{Pros}tate \underline{Reg}istration \underline{Net}work) pipeline to register T2-w MRI and histopathology images, which consists of image preprocessing, transformation estimation by deep neural networks, and postprocessing, as shown in Figure 1.
\begin{figure*}[!hbt]
	\centering
	\includegraphics[width=14cm, keepaspectratio]{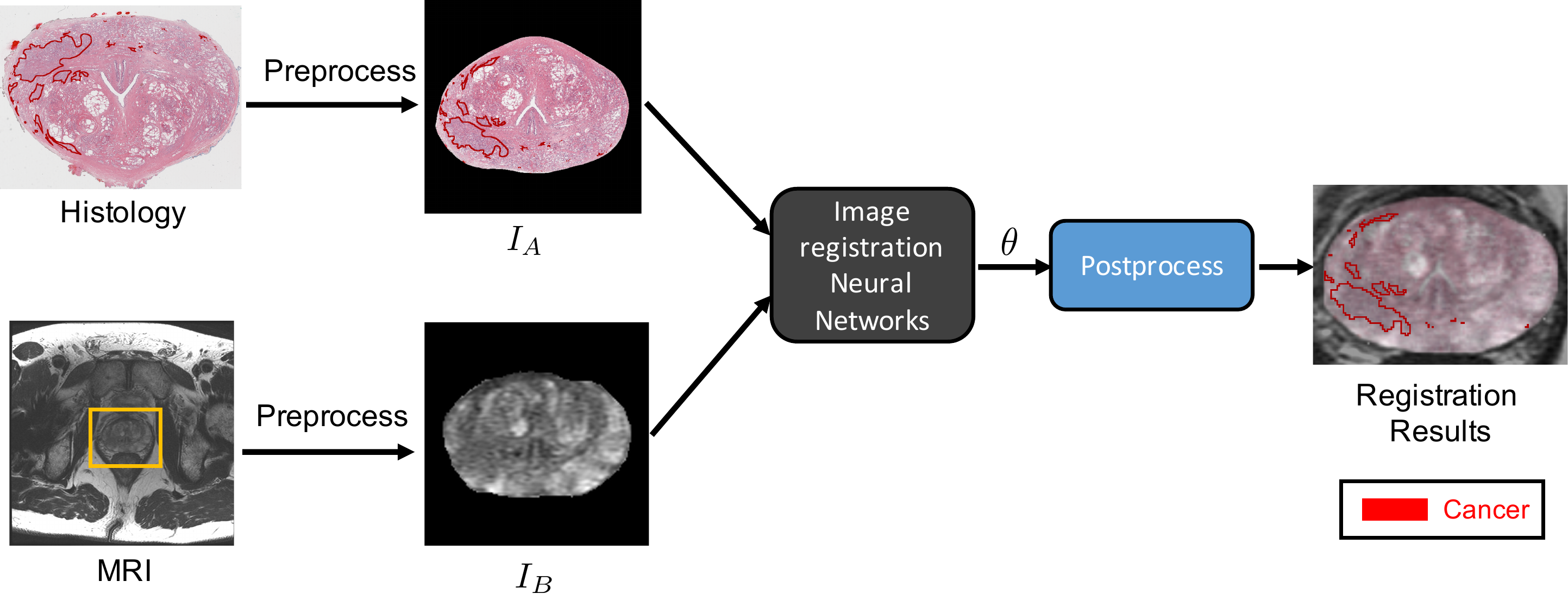}
	\caption{Proposed  pipeline for registration of MRI and histopathology images. The yellow rectangle highlights the prostate in the MRI slice.
		The preprocessed images $I_A$ and $I_B$ represent the moving and the fixed images, respectively. 
		Images $I_A$ and $I_B$ are fed into the image registration neural network to estimate  $\theta$ that represents the affine and nonrigid transformation parameters. 
		Cancer labels (the red outlines) in the histopathology slice are then deformed into the MRI slice using the estimated transformations.
	}
	\label{fig:pipeline}
\end{figure*}

\subsubsection{Preprocessing}
Mounting of tissue sections on glass slides can produce several significant artifacts, including tissue shrinkage, in-plane rotation and horizontal flipping, that will affect alignment with the corresponding MR images. 
We manually corrected for the gross rotation angle and determined whether horizontal flipping was present for each histopathology slice, as shown in $I_A$ in Figure 1. We applied the same rotation and flip transformations to the binary mask of the prostate, cancer regions, urethra, and other regions of the prostate in the histopathology slice.
A bounding box around the prostate mask was applied to extract prostate slices from the T2-w MRI, as shown in $I_B$ in Figure~\ref{fig:pipeline}.
We normalized the intensity of each cropped MRI slice from 0 to 255.
The histopathology and MRI images were multiplied by the corresponding prostate masks to facilitate the registration process.
The resulting images $I_A$ and $I_B$ were then resampled to $240\times240$ before feeding into the registration neural networks.
This preprocessing procedure has been applied to images going through the CNNGeometric and ProsRegNet networks.

\subsubsection{Image Registration Neural Networks}
\begin{figure*}[!hbt]
	\centering
	\includegraphics[width=12cm,keepaspectratio]{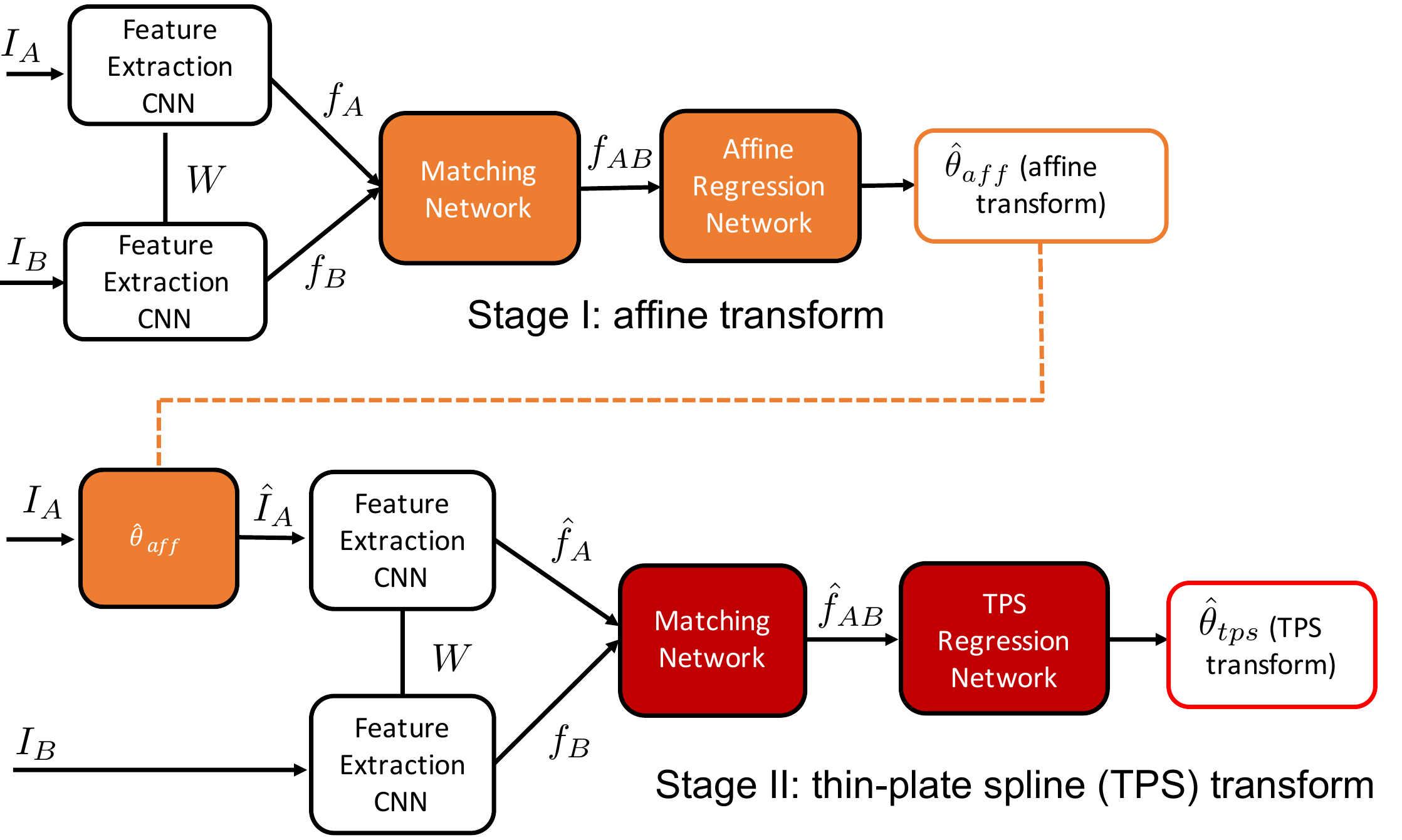}
	\caption{Two-stage registration framework using deep neural networks~\cite{Rocco17}.
		The first stage estimates an affine transform that globally aligns the two images. The second stage uses the affine transform as initialization to determine a thin-plate spline transform. Composing the two transforms gives the resulting correspondence map between $I_A$ and $I_B$. }
	\label{fig:stage12}
\end{figure*}

Both ProsRegNet and CNNGeometric registration networks consisted of feature extraction, feature matching, and transformation parameter estimation and utilized a two-stage registration architecture (see Figure~\ref{fig:stage12}). 
In the first stage, an affine transformation was estimated to align the two images globally.
In the second stage, the affine transformation is used as an initial transform to facilitate the estimation of a more accurate thin-plate spline (TPS) transformation. 
There are two major differences between our ProgRegNet model and the prior CNNGeometric model. First, our ProgRegNet model used image intensity differences to train the registration networks in an unsupervised manner, while CNNGeometric used a loss based on point location differences in a supervised training. Second, our ProsRegNet model improved the stability of the training by parameterizing the transformations using the sum of an identity transform and the estimated parameter vector scaled by a small weight, while CNNGeometric directly used the estimated parameter vector.

\begin{figure}[!hbt]
	\centering
	\includegraphics[width=9cm,keepaspectratio]{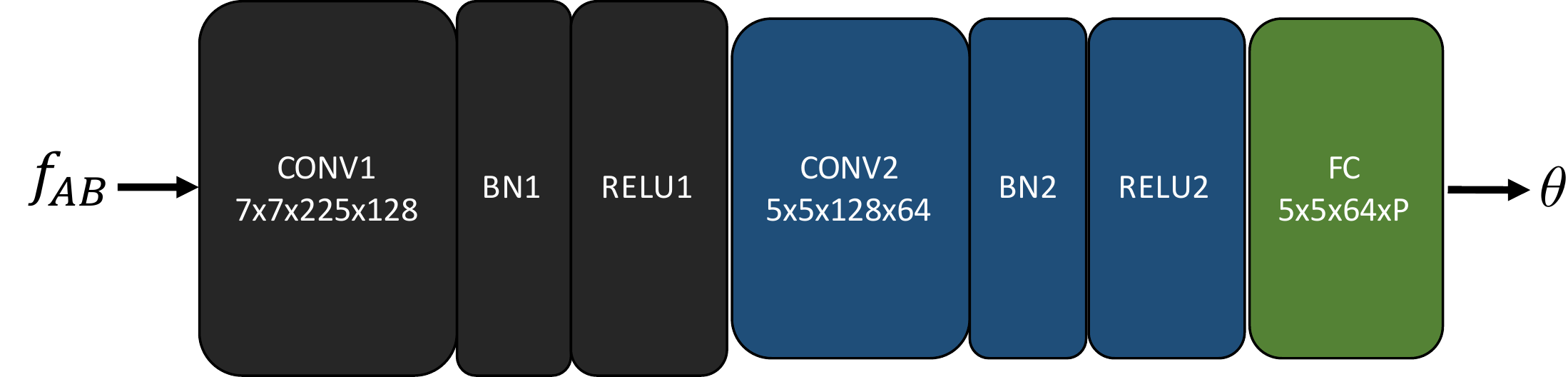}
	\caption{Regression network for estimating transformation parameters from the correspondence map $f_{AB}$~\cite{Rocco17}.}
	\label{fig:regression}
\end{figure}

We use the same feature extraction and regression networks as in~\cite{Rocco17}.
The inputs to the geometric matching networks are a moving image $I_A$ and a fixed image $I_B$.
Those two images were passed through the same pre-trained feature extraction convolutional neural network (ResNet-101~\cite{HeKaiming2016} network cropped at the third layer)
to produce the corresponding feature maps $f_A$ and $f_B$, respectively.
Each feature map is an image of size $(w,h)$ whose value at each voxel is a $d$-dimensional vector, where $d$ is the number of features.
The feature maps $f_A$ and $f_B$ were fed into a correlation layer followed by normalization.
The correlation layer combines $f_A$ and $f_B$ into a single correlation map $c_{AB}$ of the same size.
At each voxel location $(i,j)$, $c_{AB}(i,j)$ is a vector of length $wh$ whose $k$-th element is given by:
\begin{equation}
c_{AB}(i, j, k) = f_B(i,j)^T f_A(i_k,j_k)
\end{equation}
where $k = h(j_k - 1) + i_k$.
The correlation map $c_{AB}$ was normalized using a rectified linear unit (ReLU) followed by a channel-wise $L^2$-normalization.
The resulting tentative correspondence map $f_{AB}$
was passed through a regression network to estimate parameters of the geometric transformation between $I_A$ and $I_B$. 
The regression network consisted of two stacked layers, where each layer begins with a convolutional unit and is followed by batch normalization and ReLU. A final fully connected (FC) layer regresses the parameters of the geometric transform, as shown in Figure~\ref{fig:regression}. 

The output of the regression network ($\theta$) is a vector of 6 elements when performing affine registration.
Unlike~\cite{Rocco17} that directly use $\theta = (\theta_1,\cdots,\theta_6)$ as the affine matrix, we propose to use $\alpha\theta +  \theta_{Id}^{aff}$, where $\alpha$ is a small number and $\theta_{Id}^{aff}$ is the parameter vector for identity affine transform.
To be more specific, the affine transformation associated with $\theta$ is given by:
\begin{equation}
\phi_{\theta}(x,y) = \begin{bmatrix}
1 + \alpha \theta_1 & \alpha \theta_2 \\ \alpha \theta_4 & 1 + \alpha \theta_5
\end{bmatrix} \begin{bmatrix} x \\y \end{bmatrix} + \begin{bmatrix} \alpha \theta_3 \\ \alpha \theta_6 \end{bmatrix}
\end{equation}
where $(x,y)$ is any spatial location, and we choose $\alpha = 0.1$ in this paper.

Using $\alpha\theta + \theta_{Id}^{aff}$ instead of $\theta$ as the affine matrix guarantees that the initial estimate of $\phi_{\theta}$ during the network training is close to the identity map and thus improves the stability of our registration network.
We parameterize the nonrigid transformations using a thin-plate spline grid of size 6x6 instead of $3\times3$ in~\cite{Rocco17} for more accurate registration.
This requires $\theta$ to be a vector of $2\times 6 \times6 = 72$ elements.
Similarly, we use $\alpha\theta + \theta_{Id}^{tps}$ instead of $\theta$ to parameterize the nonrigid transforms, where $\theta_{Id}^{tps}$ is the parameter vector for the identity thin-plate spline transform.

Unlike~\cite{Rocco17} that uses the differences between the original and deformed coordinate locations (location matching error) as the loss function, we define the loss function as the sum of squared differences (SSD) between the fixed and the deformed image (since $I_A$ and $I_B$ are from the same modality during the training):
\begin{equation}
loss(\theta) = \sum_{i = 1}^{W}\sum_{j = 1}^{H}|| I_A(i,j) - I_B\circ\phi_{\theta}(i,j)||^2
\end{equation}
where $\phi_{\theta}$ is the transformation parameterized by $\theta$, and W and H are the width and height of the images.
Since all MR and histopathology images have been masked during the preprocessing, our SSD cost can quickly drive the registration process during the training.

\subsubsection{Postprocessing}
After affine and deformable image registrations, the histopathology images and the prostate, urethra, anatomic landmarks, and cancer labels on the histopathology images were mapped to the corresponding MRI slices using the estimated composite (affine + deformable) transformation.
Although the histopathology images were resampled to have a size of $240\times240$, the deformed histopathology images still have the same size as the original histopathology images since we applied the estimated composite transformation directly to the original high-resolution histopathology images.
For visualization purposes, sampling artifacts in the deformed images were removed by binary thresholding to set the intensity of pixels outside the prostate to be zero.

\subsection{Training Dataset}
\begin{figure}[!hbt]
	\centering
	\begin{subfigure}[t]{0.5\textwidth}
		\centering
		\includegraphics[width=7cm,keepaspectratio]{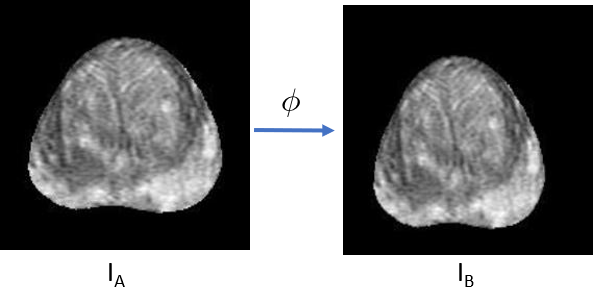}
		\caption{Synthetic MR image pair}
	\end{subfigure}
	\hspace{0.025in}
	\begin{subfigure}[t]{0.5\textwidth}
		\centering
		\includegraphics[width=7cm,keepaspectratio]{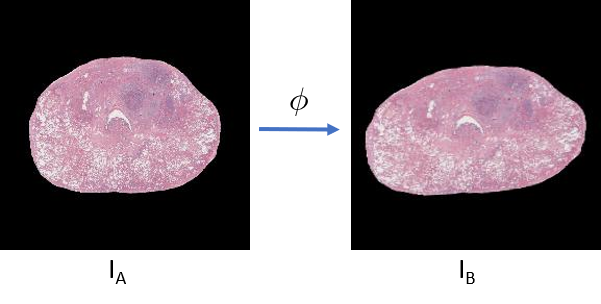}
		\caption{Synthetic histopathology image pair}
	\end{subfigure}
	\caption{Generating training dataset by applying known transformations. $I_A$ is the original image, $\phi$ is either an affine or thin-plate spline transform, and $I_B$ is the deformed image by applying $\phi$ to $I_A$. Each tuple $(I_A,I_B,\phi)$ is considered as one training example.}
	\label{fig:training_data}
\end{figure}

Since the ground truth spatial correspondences between the MRI and histopathology images are lacking,
we trained our neural networks using uni-modal image pairs generated by synthetic transformations (Figure~\ref{fig:training_data}).
For each 2D image $I_A$, we applied a simulated transformation $\phi$ to deform it into the image $I_B$.
The 3-tuple $(I_A,I_B, \phi)$ will be used as one training example.
The transformation $\phi$ can be either an affine transform or a thin-plate spline transform.
To guarantee the plausibility of the simulated transformations, the variables used to parameterize the transformations were randomly sampled from bounded intervals.
When simulating the affine transformations, the rotation angle ranged from -10 degrees to +10 degrees, the scaling coefficients ranged from 0.8 to 1.2, the shifting coefficients were within 5\% of the image size, and the shearing coefficients were within 5\%.
When simulating the thin-plate-spline transformations, the movement of each control point was within 5\% of the image size.
We chose these intervals as they represent typical transformation ranges we observed when using RAPSODI and were shown to be sufficiently wide to cover the transformations observed in our diverse patient cohorts.
For the training, we used 1,390 MRI and histopathology images and the corresponding prostate masks of 99 patients from Cohort 1.
Prostate masks were used to train the affine registration network and masked MRI and histopathology images were used to train the deformable registration network.
Although our registration neural network was trained with image pairs of the same modality, we will show that it can be generalized to the multi-modal registration of MRI and histopathology images for all three cohorts.

\subsection{Experiments}
\label{sec:exp}
We trained the neural networks on the NVIDIA GeForce RTX 2080 GPU (8GB memory, 14000 MHz clock speed).
We used an initial learning rate of 0.001, a learning rate decay of 0.95, a batch size of 64, and the Adam optimizer~\cite{KingmaDiederik2017AAMf}, for which both the affine and deformable registration networks were trained with 50 epochs. For each deformation model, the network with the minimum validation loss during the training was used in the testing.

In total, we experimented with three different approaches for registration of MRI and the corresponding histopathology images: the traditional RAPSODI registration framework~\cite{rusu2020registration} (RAPSODI), a prior deep learning registration framework developed by Rocco et al.~\cite{Rocco17} (CNNGeometric), and our deep learning ProsRegNet pipeline (ProsRegNet), 
We tested the RAPSODI approach on the Intel Core i9-9900K CPU (8-Core, 16-Thread, 3.6 GHz (5.0 GHz Turbo)) and tested the CNNGeometric and ProsRegNet approaches on the GeForce RTX 2080 GPU.
In total, we used datasets of 53 prostate cancer patients (12 from Cohort 1, 16 from Cohort 2, and 25 from Cohort 3) to evaluate the performance of the above three registration approaches.

\subsection{Evaluation Metrics} 
The Dice coefficient, the Hausdorff distance, and the mean landmark error were used to evaluate the alignment accuracy for the deformed histopathology and the corresponding MRI images.
The Dice coefficient measures the relative overlap between $M_A$ and $M_B$, which is given by:
\begin{equation}
C_{D} = \frac{2|M_A \cap M_B|}{|M_A| + |M_B|}
\end{equation}
where $M_A$ denotes the deformed histopathology prostate mask, $M_B$ denotes corresponding MRI prostate mask, and $|\cdot|$ denotes the cardinality (number of elements) of a set.

The Hausdorff distance measures how close the prostate boundaries are defined in $A$ and $B$, which is given by:
\begin{equation}
d_{H} = \max\left\{ \sup_{a\in M_A} \inf_{b\in M_B} ||a - b||,\sup_{b\in M_B} \inf_{a\in M_A}||a - b||\right\}
\end{equation}
where $||\cdot||$ is the standard $L^2$ metric, $sup$ represents the supremum, and $inf$ represents the infimum.

The mean landmark error measures the accuracy of point-to-point correspondences found by image registration.
Let $\phi$ denote the resulting transformation from image registration.
Our experts labeled $N$ landmark pairs in the fixed T2-w MRI and the moving histopathology image, denoted by  $(p_1,p_1'),\cdots,(p_N,p_N')$.
Then the mean landmark error for the resulting transformation $\phi$ from image registration is given by: 
\begin{equation}
d_L = \frac{1}{N}\sum_{i=1}^{N}||p'_i - \phi(p_i)||
\label{eq:lmk_error}
\end{equation}

We used an identical approach to evaluate the
distance between urethra segmentation on MRI and the corresponding deformed urethra segmentation on histopathology images (urethra deviations).
All evaluation measures were computed on a slice by slice basis in 2D and averaged across several slices to obtain per patient measures.

\section{Results}
\label{sec:results}

\begin{figure}[!hbt]
	\centering
	\includegraphics[width=0.48\textwidth,keepaspectratio]{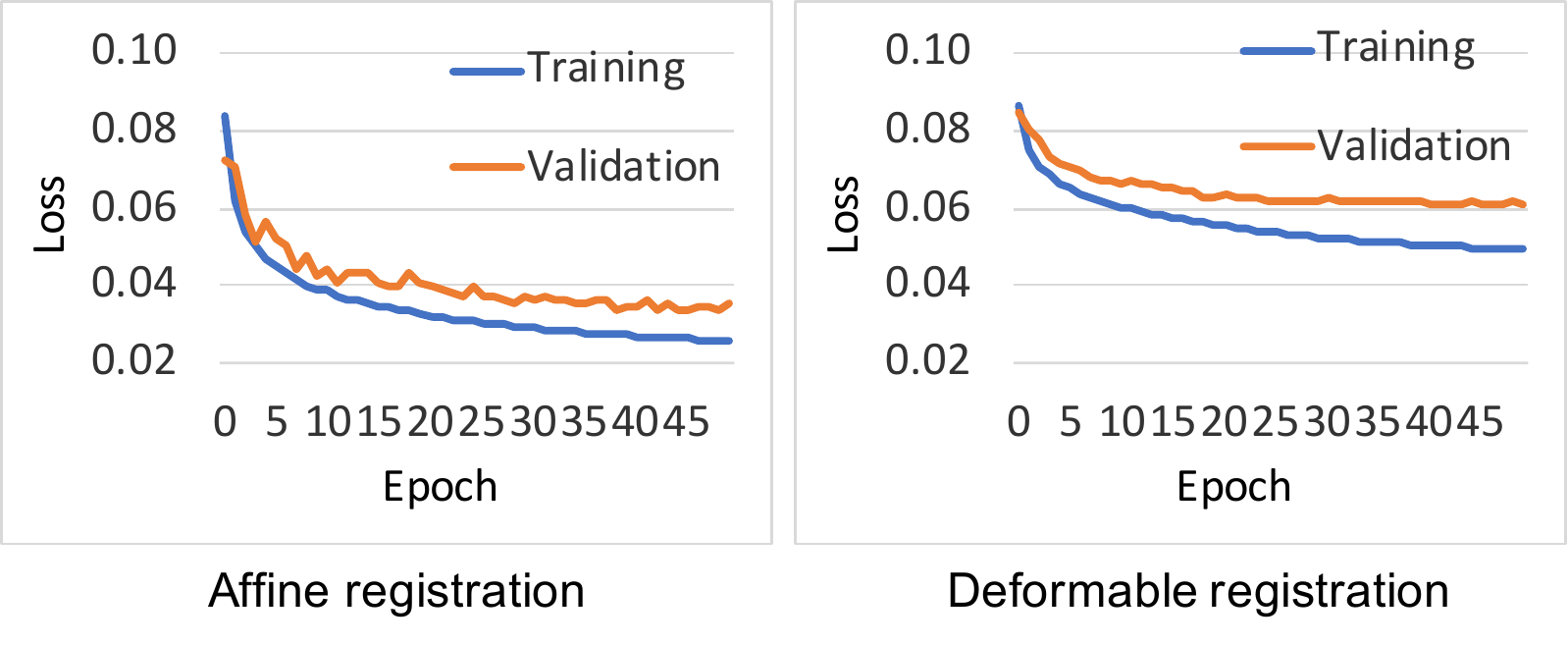}
	\caption{Training loss and validation loss curves of ProsRegNet affine and deformable registration networks.}
	\label{fig:learning_curve}
\end{figure}

Figure~\ref{fig:learning_curve} shows the training loss and validation loss curves of the ProsRegNet affine and deformable registration networks. From this figure, we can see that the validation loss has converged at 50 epochs for both networks and there is no issue of overfitting. We also notice that we had a slight unrepresentative sample for the training and we except better performance when the networks were trained with a large dataset.

\begin{figure}[!hbt]
	\centering
	\includegraphics[width=0.48\textwidth,keepaspectratio]{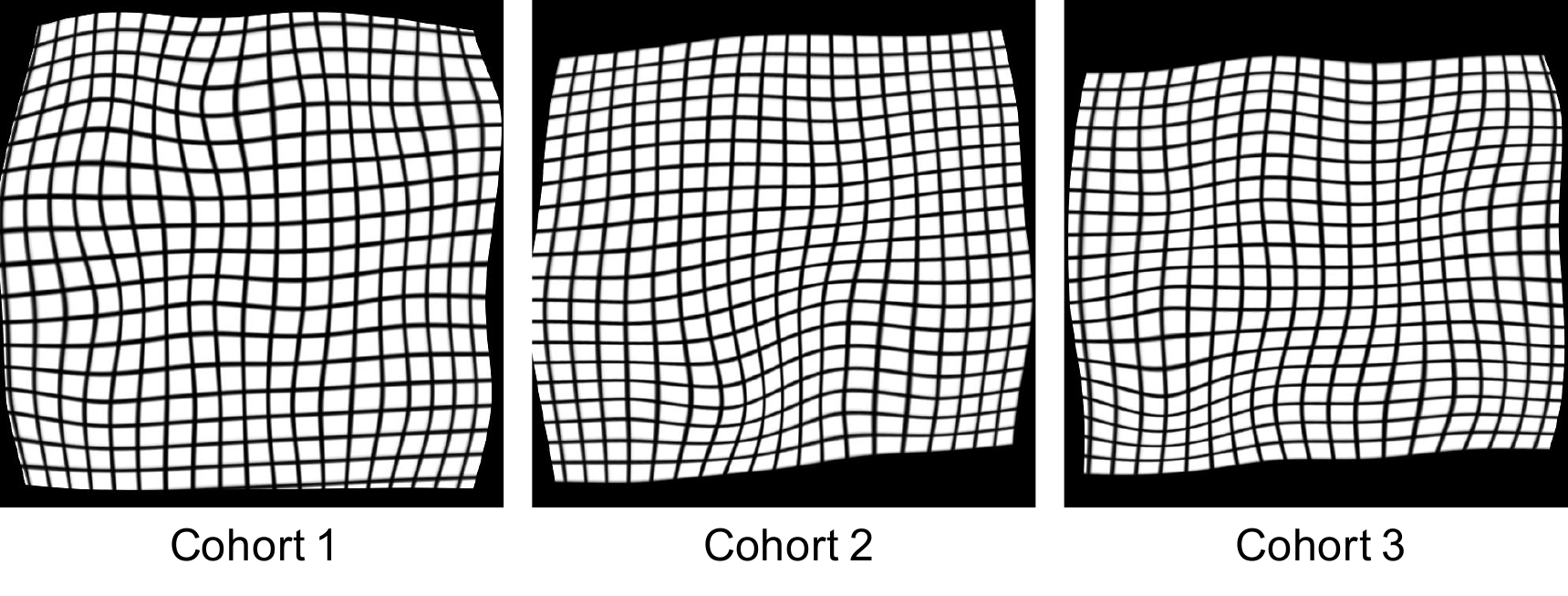}
	\caption{Typical deformed grid images from ProsRegNet registration.}
	\label{fig:grid}
\end{figure}

To evaluate the plausibility of the estimated geometric transformations, we use each of them to deform a 2D grid image.
	By investigating all deformed grid images, we conclude that the composite transformations estimated by our ProsRegNet network are smooth and biologically plausible.
Figure~\ref{fig:grid} shows a typical deformed grid image from each cohort.

\subsection{Qualitative Alignment Accuracy}
\begin{figure*}[!hbt]
	\centering
	\includegraphics[width=14cm,keepaspectratio]{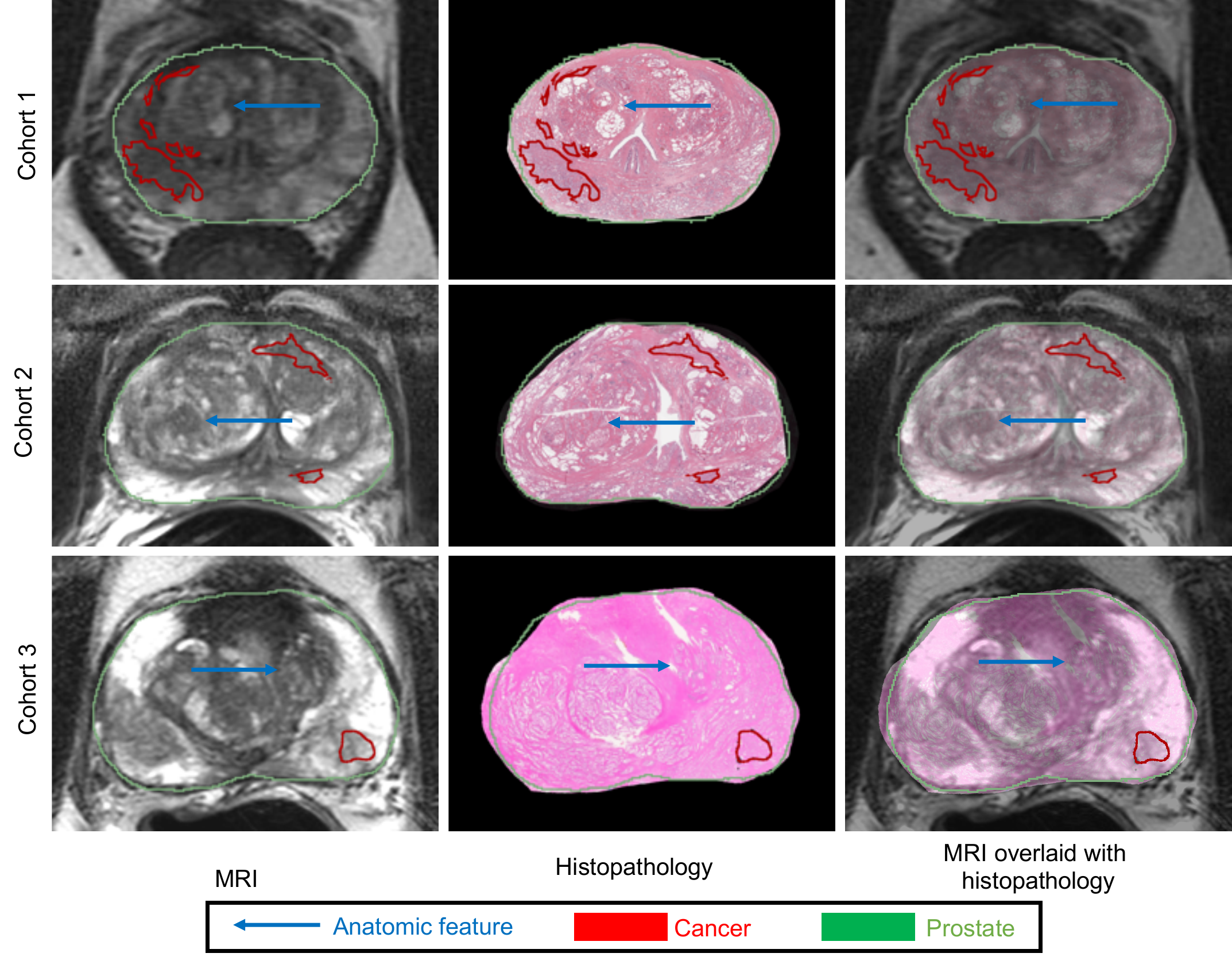}
	\caption{Registration results for three different subjects (one from each cohort) using the proposed ProsRegNet deep learning registration pipeline. 
		The MRI slices were chosen as the fixed images.
		(Left) MRI, (Middle) registered histopathology image, (Right) MRI overlaid with registered histopathology image. Cancer labels from the histopathology images were mapped onto MRI using estimated transformations from image registration.}
	\label{fig:reg_results}
\end{figure*}

Figure~\ref{fig:reg_results} shows the registration results of three patients with large cancerous regions (one from each cohort).
The prostate boundaries on the MRI and the histopathology sections appeared well aligned for all three subjects, suggesting that the ProsRegNet pipeline achieved accurate global alignment of the prostate.
Anatomic regions of the prostate on the MR and the histopathology images were also well aligned. Accurate alignment of anatomic regions indicates that the ProsRegNet pipeline has achieved promising alignment of local prostate features.
The results in Fig.~\ref{fig:reg_results} demonstrate that our ProsRegNet pipeline generalizes across cohorts even if they were not part of the training, showing accurate registration for images from different cohorts acquired by different protocols.
Our accurate alignment of the histopathology and MRI images suggests that we can carefully map the cancer labels in the histopathology images to the corresponding MRI slices using the estimated transformations. 

\subsection{Quantitative Results}
\begin{figure*}[!hbt]
	\centering
	\includegraphics[width=18cm,keepaspectratio]{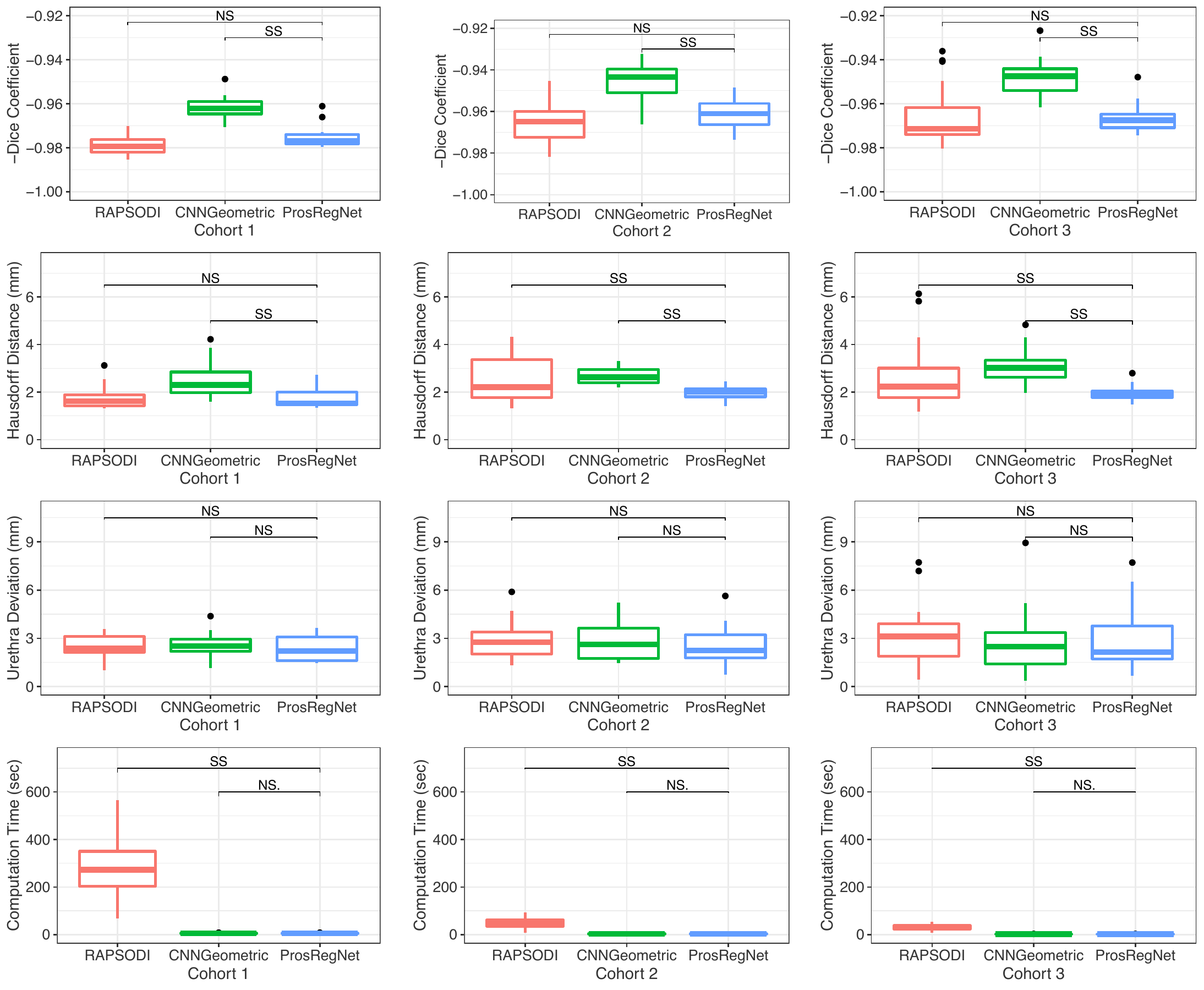}
	\caption{Box plots of different measures for the RAPSODI, CNNGeometric, and ProRegNet registration approaches of three cohorts. 
		SS: statistically significant ($p\le0.05$), NS: not significant ($p > 0.05$).
	}
	\label{fig:comparision}
\end{figure*}

We evaluated various measures to assess the quality of alignment between the histopathology images and corresponding MRI slices. Those measures assess the overall alignment of the prostate (Dice coefficient), the distance between the prostate boundaries (Hausdorff Distance), and anatomic landmark deviation.
Moreover, we also evaluated the execution time of the RAPSODI, CNNGeometric, and ProsRegNet approaches.
Figure~\ref{fig:comparision} shows the box plots of the Dice Coefficient, Hausdorff distance, urethra deviation, and computation time of different approaches for all three cohorts.
The results show that there is no significant difference (p-value$>$0.05) between the Dice coefficient and the urethra deviation of the RAPSODI and ProsRegNet approaches for all three cohorts.
Our ProsRegNet approach achieved significantly lower ($p\le0.05$) Hausdorff distance than the RAPSODI approach for the second and the third cohorts.
Our ProsRegNet approach has achieved significantly higher Dice coefficient and lower Hausdorff distance than the deep learning CNNGeometric approach for all three cohorts.
Also, there is no significant difference between the urethra deviation of all three approaches for all cohorts.
Notice that both our ProsRegNet and the CNNGeometric deep learning approaches were at least 20x faster to register the images than the iterative optimization performed by RAPSODI.
In summary, the ProsRegNet pipeline has achieved better alignment near the prostate boundary than the RAPSODI approach while being several orders of magnitude faster, and it has also achieved better alignment of the overall shape and boundary of the prostate than prior CNNGeometric deep learning approach.

Table~\ref{table:stats} summarizes the Dice coefficients for the whole prostate, Hausdorff distances for the prostate boundary, urethra deviations, and anatomic landmark errors after registration for the three cohorts. 
The results show that both the ProsRegNet and RAPSODI approaches have achieved a higher Dice Coefficient than the prior CNNGeometric approach.
The high Dice coefficient indicates that our ProsRegNet pipeline can accurately align the overall shape and edges of the prostate for all of the three cohorts.
The results also show that the ProsRegNet pipeline achieved a lower Hausdorff distance than both the RAPSODI and CNNGeometric approaches.
The low Hausdorff distance implies that our ProsRegNet pipeline can have a small registration error of no more than 2mm near the prostate boundary.
No significant differences were found between the RAPSODI, ProsRegNet, and CNNGeometric approaches in terms of urethra deviation and landmark error.
The urethra deviation and landmark error indicate that our ProsRegNet pipeline has an average registration error of no more than 3mm inside the prostate.
It is notable that the average running time of the ProsRegNet and CNNGeometric approaches was 1-4 seconds, compared to 31-264 seconds of the state-of-the-art RAPSODI approach and compared to running times of 120-750 seconds reported for other traditional approaches~\cite{LiLin2017Coev,LOSNEGARD201824}.

\begin{table*}[!hbt]
	\caption{Registration results of the RAPSODI, CNNGeometric, ProsRegNet approaches of three cohorts. }
	\begin{center}
		\begin{tabular}{|c|c|c|c|c|c|c|}
			\hline
			\multirow{2}{*}{Dataset} & Registration	& Dice & Hausdorff& Urethra & Landmark & Computation \\
			& Approach & Coefficient  & Distance (mm) & Deviation (mm) & Error (mm) & Time (second) \\
			[0.5ex] 
			\hline
			\hline
			\multirow{2}{*}{Cohort 1}& RAPSODI & \textbf{0.979 ($\pm$ 0.01)} & 1.83 ($\pm$ 0.50) & 2.48 ($\pm$ 0.78)& 2.88 ($\pm$ 0.73) & 264 ($\pm$ 150)\\
			& CNNGeometric & 0.962 ($\pm$ 0.01) & 2.43 ($\pm$ 0.83) & 2.62 ($\pm$ 0.86) & 2.72 ($\pm$ 0.75) & \textbf{4 ($\pm$2)} \\
			&ProsRegNet  & 0.975 ($\pm$ 0.01) & \textbf{1.72 ($\pm$ 0.42)} & \textbf{2.37 ($\pm$ 0.76)} & \textbf{2.68 ($\pm$ 0.68)} & \textbf{4 ($\pm$2)} \\ 
			\hline
			\hline
			\multirow{2}{*}{Cohort 2}& RAPSODI & \textbf{0.965 ($\pm$ 0.01)} & 2.58 ($\pm$ 1.05) & 2.96 ($\pm$ 1.23)& NA & 60 ($\pm$ 47)\\
			& CNNGeometric & 0.948 ($\pm$ 0.01) & 3.05 ($\pm$ 0.69) & 2.78 ($\pm$ 2.03) & NA & \textbf{3 ($\pm$1)} \\
			&ProsRegNet  & 0.961 ($\pm$ 0.01) & \textbf{1.98 ($\pm$ 0.28)} & \textbf{2.51 ($\pm$ 0.82)} & NA & \textbf{3 ($\pm$1)} \\ 
			\hline
			\hline
			\multirow{2}{*}{Cohort 3}& RAPSODI & 0.966 ($\pm$ 0.01) & 2.62 ($\pm$ 1.32) & 3.3 ($\pm$ 1.90)& NA  & 31 ($\pm$ 11)\\
			&CNNGeometric  & 0.946 ($\pm$ 0.01) & 2.68 ($\pm$ 0.33) & \textbf{2.83 ($\pm$ 1.2)} & NA & \textbf{1 ($\pm$1)} \\
			&ProsRegNet  & \textbf{0.967 ($\pm$ 0.01)} & \textbf{1.96 ($\pm$ 0.29)} & 2.91 ($\pm$ 1.99) & NA & \textbf{1 ($\pm$1)} \\ 
			\hline
		\end{tabular}
	\end{center}
	\label{table:stats}
\end{table*}

\subsection{Alignment of Prostate Cancers}
One major goal of MRI-histopathology registration is 
	to map the ground truth cancer labels from the histopathology images onto MRI.
	Here, we evaluate the accuracy of different approaches for registering cancerous regions using patients from first cohort and the second cohorts.
	For the first cohort, two body imaging radiologists with more than four years of experience manually labeled regions of clinically significant prostate cancer on T2-w MRI of 35 patients.
	The following exclusion criterion was applied to handle inconsistency between the radiologists'  and pathologists' annotations: (1) the size of two cancer labels of the same region differs by more than 100\%, (2) there is no overlap between two cancer labels of the same region, (3) cancer labels are too tiny (less than 25 pixels).
	For the second cohort, the authors of the dataset have provided cancer labels on MRI by performing landmark-based registration of MRI and histopathology images.
	Table~\ref{table:cancer_alignment} 
shows the Dice coefficient and Hausdorff distance between cancer labels from the radiologists' or landmark-based registration  and cancer labels achieved by each of the registration approaches.
	The results show that ProsRegNet achieved better alignment of the prostate cancer boundaries (Hausdorff distance) than RAPSODI and CNNGeometric for both cohorts.
	Although CNNGeometric achieved slightly higher Dice coefficient than RAPSODI and ProsRegNet for the second cohort, our ProsRegNet approach achieved the highest Dice coefficient for the first cohort.
	In summary, our ProsRegNet approach has achieved comparable or better alignments of cancerous regions relative to CNNGeometric and RAPSODI.
	Notice that the accuracy of our analysis may be compromised by inconsistency between the radiologist's cancer labels and the pathologists' cancer labels (first cohort), and also errors in landmark-based registration (second cohort).

\begin{table}[!hbt]
	\caption{Accuracy of the RAPSODI, CNNGeometric, ProsRegNet approaches for aligning cancerous regions.}
	\small
	\begin{center}
		\begin{tabular}{|c|c|c|c|}
			\hline
			\multirow{2}{*}{Dataset} & Registration	& Dice & Hausdorff \\
			& Approach & Coefficient  & Distance (mm)  \\
			[0.5ex] 
			\hline
			\hline
			\multirow{2}{*}{Cohort 1} & RAPSODI & 0.624 ($\pm$ 0.12) & 6.02 ($\pm$ 2.78)  \\
			& CNNGeometric & 0.610 ($\pm$ 0.11) &  5.70 ($\pm$ 2.22) \\
			&ProsRegNet  &  \textbf{0.628 ($\pm$ 0.10)} & \textbf{5.42 ($\pm$ 2.32)}  \\ 
			\hline
			\hline
			\multirow{2}{*}{Cohort 2}& RAPSODI & 0.573 ($\pm$ 0.13) & 5.42 ($\pm$ 2.00) \\
			& CNNGeometric & \textbf{0.575 ($\pm$ 0.12)} & 5.34 ($\pm$ 2.14)  \\
			&ProsRegNet  & 0.563 ($\pm$ 0.14) & \textbf{4.87 ($\pm$ 1.53)}  \\ 
			\hline
			\hline
		\end{tabular}
	\end{center}
	\label{table:cancer_alignment}
\end{table}

\subsection{Other Training Schemes}
In this section, we investigate two additional training schemes, one ProsRegNet and the other one for CNNGeometric.
For the first training scheme, we trained both the affine and deformable registration  networks of ProsRegNet directly by the prostate masks of 99 patients from the first cohort and tested the performance on 53 patients from three cohorts (see Table~\ref{table:other}). 
Compared to results presented in Table~\ref{table:stats}, 
training and testing ProsRegNet with only prostate masks has improved the alignment of prostate boundaries, with Dice coefficient increased by 0.4\%-1.0\% and Hausdorff distance decreased by 13.4\%-18.7\%.
However, this training scheme has also deteriorated the registration results inside the prostate, with urethra deviation increased by 13.5\%-25.7\% and landmark error increased by 24.6\%.
Those results show that training the ProsRegNet model with masked MRI and histopathology images facilitates the alignment of features inside the prostate.
Since alignment of features inside the prostate is more important than alignment of the prostate boundaries, we do not recommend training and testing ProsRegNet only on the prostate masks.

\begin{table*}[!hbt]
	\caption{Registration results of ProsRegNet trained with only prostate masks and CNNGeometric trained with multi-modal image pairs. }
	\begin{center}
		\begin{tabular}{|c|c|c|c|c|c|}
			\hline
			\multirow{2}{*}{Dataset} & Registration	& Dice & Hausdorff& Urethra & Landmark \\
			& Approach & Coefficient  & Distance (mm) & Deviation (mm) & Error (mm) \\
			[0.5ex] 
			\hline
			\hline
			\multirow{2}{*}{Cohort 1}& ProsRegNet (masks only) & 0.979 ($\pm$ 0.01) & 1.49 ($\pm$ 0.44) & 2.98 ($\pm$ 0.82)& 3.39 ($\pm$ 0.68)\\
			& CNNGeometric (multi-modal) & 0.960 ($\pm$ 0.01) & 2.42 ($\pm$ 0.55) & 2.55 ($\pm$ 0.73)& 2.79 ($\pm$ 0.74) \\
			\hline
			\hline
			\multirow{2}{*}{Cohort 2}& ProsRegNet (masks only) & 0.971 ($\pm$ 0.01) & 1.61 ($\pm$ 0.33) & 2.85 ($\pm$ 1.34)& NA\\
			& CNNGeometric (multi-modal) & 0.910 ($\pm$ 0.03) & 4.08 ($\pm$ 1.14) & 2.82 ($\pm$ 1.34) & NA\\
			\hline
			\hline
			\multirow{2}{*}{Cohort 3} & ProsRegNet (masks only) & 0.976 ($\pm$ 0.01) & 1.60 ($\pm$ 0.38) & 3.57 ($\pm$ 2.28)& NA\\
			& CNNGeometric (multi-modal) & 0.947 ($\pm$ 0.01) & 3.00 ($\pm$ 0.82) & 3.17 ($\pm$ 2.07) & NA \\
			\hline
		\end{tabular}
	\end{center}
	\label{table:other}
\end{table*}

For the second training scheme, we investigated the efficacy of training a multi-modal deep learning network on MRI-histopathology image pairs pre-aligned by RAPSODI.
We chose the CNNGeometric model over the ProsRegNet model since the SSD loss function used by the ProsRegNet model cannot be directly used for multi-modal registration.
Again, we trained  the CNNGeometric model on MRI-histopathology image pairs of 99 patients from the first cohort and evaluated its performance using 53 patients from three cohorts (see Table~\ref{table:other}).
shows the registration results of the multi-modal CNNGeometric network for the three cohorts.
Compared to results in Table~\ref{table:stats}, the performance of the multi-modal CNNGeometric model is worse than the uni-modal ProsRegNet model for both the global and local alignment of the MRI and histopathology images.
One factor that compromised the performance of the multi-modal CNNGeometric is that the MRI-histopathology image pairs used in the training are from RAPSODI registration and therefore do not have perfect spatial correspondences.

\section{Discussion}
Accurately aligning MRI with histopathology images provides a detailed answer key regarding precise cancer locations on MRI.
As such, it has tremendous potential for improving the interpretation of prostate MRI and providing labeled imaging data to establish and validate prostate cancer detection models based on radiomics or machine learning methods~\cite{MetzgerGregoryJ2016DoPC}. 
In this paper, we have developed the novel ProsRegNet deep learning approach for 2D registration of MRI and histopathology images.
It is challenging to directly train a multi-modal network for registering MR and histopathology images due to the lack of either an effective loss function for unsupervised learning or MRI-histopathology image pairs with accurate spatial correspondences for supervised learning. We tackled this problem from a different perspective by training a uni-modal ProsRegNet network which learns how to combine high-level features in the MRI and histopathology images to solve image registration problems. The trained ProsRegNet network has the capabilities to solve uni-modal registration problems in the context of MRI and histopathology images and thus can be used to register the two modalities in a multi-modal manner. Our experiments and results provide empirical evidence that, although our ProsRegNet was trained with pairs of images from the same modality, it can be generalized to achieve very accurate MRI-histopathology registration.
This paper is the first attempt to apply deep learning to the registration of MRI and histopathology images of the prostate.

Our study is the largest prostate MRI-histopathology registration study, using  654 of pairs of histopathology and MRI slices of 152 prostate cancer patients from three different institutions and MRIs from three different manufactures.
The wide range of parameters of synthetic transformations used during the training allowed ProgRegNet to accurately recover large affine and deformable transformations observed in three different cohorts, which include MR images acquired with or without using an endorectal coil, as well as histopathology images acquired as whole mounts, quadrants or at low resolution.
We showed that our ProsRegNet pipeline achieved a very high Dice coefficient (0.96-0.98), a very low Hausdorff distance (1.7-2.0mm), a relatively low urethra deviation (2.4-2.9mm) and a relatively low landmark error (2.7mm) compared to results reported in previous studies~\cite{ChappelowJonathan2011Erom,KalavaguntaChaitanya2015Roiv,LiLin2017Coev,LOSNEGARD201824,ParkHyunjin2008RMfH,ReynoldsH.M.2015Doar,WardAaronD2012Prod,WuHoldenH.2019Asup}. 
Moreover, in a direct comparison of the state-of-the-art RAPSODI pipeline~\cite{rusu2020registration}, we showed that ProsRegNet achieved slightly better performance while being 20x-60x faster.
This allows our ProsRegNet approach to execute the histopathology-MRI registration in real-time interactive software, otherwise not possible with any previous method.
By significantly speeding up the registration process, our approach can help to create a large dataset of labeled MRI using ground-truth histopathology images which is crucial for the training of prostate cancer detection methods on pre-operative MRI using machine learning. 

Even recent deep learning cancer prediction studies~\cite{CaoRuiming2019JPCD,SumathipalaYohan2018Pcdf} that use histopathology images as the reference, rely on cognitive alignment (mentally projecting the histopathology images onto the MRI) to create cancer labels on MRI. 
This time-consuming labeling is inaccurate and biases the labels towards visible extent of cancer on MRI (known to underestimate the real size of cancer~\cite{PiertMorand2018Aots} and missing MRI-invisible lesions). 
Our ProsRegNet pipeline allows the efficient creation of labels on MRI with accurate borders, including MRI invisible lesions.
In addition, once trained, our deep learning network is parameter-free when registering unseen pairs of MRI and histopathology images, alleviating the need of modifying registration hyperparameters, e.g. step size, number of iterations.
By making the registration set up less complicated, our approach is more accessible to non-expert users than the traditional methods.

Although this study demonstrates promising results for MRI-histopathology registration, there are some limitations related to human input: prostate segmentation on MRI  and histopathology images, gross rotation and flip of the histopathology images and identifying slice-to-slice correspondences.  
Our team is working on developing methods to  automate these steps, yet they are beyond the scope of the current study. Nonetheless, our proposed work simplifies the registration step without requiring manual picking of landmarks or complex selection of features for multi-feature scoring functions~\cite{ChappelowJonathan2011Erom,LiLin2017Coev,WardAaronD2012Prod}.

We have shown that our deep learning pipeline can achieve fast and accurate registration of the histopathology and MRI images.
Accurate registration could improve radiologists’ interpretation of MRI by allowing side-by-side comparison of MR and histopathology images.
Indeed, we use these side-by-side comparisons in a multidisciplinary prostate MRI tumor board at our institution.
Accurate registration also allows mapping of the ground truth extent and grade of prostate cancer from histopathology images onto the corresponding pre-operative MRI.
Such accurate labels mapped from histopathology images on MRI will help develop and validate radiomic and machine learning approaches for detecting cancer location with the prostate based on MRI to guide biopsies and focal treatment.

\section{Conclusion}

We have developed a deep learning pipeline for efficient registration of MRI and histopathology images of the prostate for patients that underwent radical prostatectomy.
The performance of the deep neural networks for aligning the MRI and histology is promising and slightly better than state-of-the-art registration approaches. 
Compared to traditional approaches that require significant user input (e.g., careful choice of registration parameters) and considerable computing time, our pipeline achieved very accurate and efficient alignment with less user input. 
The ease of use and speed make our pipeline attractive for clinical implementation to allow direct comparison of MR and histological images to improve radiologist accuracy in reading MRI.  Furthermore, this pipeline could serve as a useful tool for image alignment in developing radiomic and deep learning approaches for early detection of prostate cancer.

\section*{Acknowledgments}
This work was supported by the Department of Radiology at Stanford University, Radiology Science Laboratory (Neuro) from the Department of Radiology at Stanford University, and the National Cancer Institute (U01CA196387 to James D. Brooks)

\bibliography{ProsRegNet}
\bibliographystyle{IEEEtran}

\end{document}